\documentclass[10pt]{wlscirep}
\usepackage[utf8]{inputenc}
\usepackage[T1]{fontenc}
\usepackage{bm}
\usepackage{siunitx}
\usepackage{xcolor}
\usepackage{amsmath,hyperref}
\usepackage{amsfonts,amssymb}
\usepackage{amsmath}
\usepackage{lineno}
\usepackage[normalem]{ulem}
\usepackage{floatrow}
\usepackage{gensymb}
\usepackage[label font=bf,labelformat=simple]{subfig}
\hypersetup{
     colorlinks=true,
     linkcolor=blue,
     filecolor=blue,
     citecolor = blue,      
     urlcolor=blue,
     }
     
     \definecolor{mygreen}{rgb}{0.0, 0.5, 0.0}
     \definecolor{myyellow}{rgb}{0.89, 0.82, 0.04}
     \definecolor{darkbrown}{rgb}{0.4, 0.26, 0.13}

\definecolor{myblue}{RGB}{81,165,231}
\usepackage{tcolorbox}
\usepackage{tabularx}
\usepackage{array}
\usepackage{colortbl,ulem}
\tcbuselibrary{skins}

\newcommand\redout{\bgroup\markoverwith
{\textcolor{red}{\rule[.5ex]{2pt}{0.4pt}}}\ULon}

\newcolumntype{Y}{>{\raggedleft\arraybackslash}X}

\tcbset{tab1/.style={fonttitle=\bfseries\large,fontupper=\normalsize\sffamily,
colback=yellow!10!white,colframe=red!75!black,colbacktitle=yellow!40!white,
coltitle=black,center title,freelance,frame code={
\foreach \n in {north east,north west,south east,south west}
{\path [fill=red!75!black] (interior.\n) circle (3mm); };},}}

\tcbset{tab2/.style={enhanced,fonttitle=\bfseries,fontupper=\normalsize\sffamily,
colback=yellow!10!white,colframe=red!50!black,colbacktitle=yellow!40!white,
coltitle=black,center title}}

\title{Topological defects reveal the plasticity of glasses}
\author[1,2,3]{Matteo Baggioli}
\affil[1]{School of Physics and Astronomy, Shanghai Jiao Tong University, Shanghai 200240, China}
\affil[2]{Wilczek Quantum Center, Shanghai Jiao Tong University, Shanghai 200240, China}
\affil[3]{Shanghai Research Center for Quantum Sciences, Shanghai 201315, China\vspace{0.1cm}}

\affil[$\,$]{$\,$ \color{blue}b.matteo@sjtu.edu.cn\color{black}}

\begin{abstract}
\textbf{Mixing theoretical topological structures with cutting-edge simulation methods, a recent study in Nature Communications has finally confirmed the existence of topological defects in glasses and their crucial role for plasticity.}
\end{abstract}
\begin{document}

\flushbottom
\maketitle
\thispagestyle{empty}

\begin{figure}[h]\centering
\includegraphics[width=\linewidth]{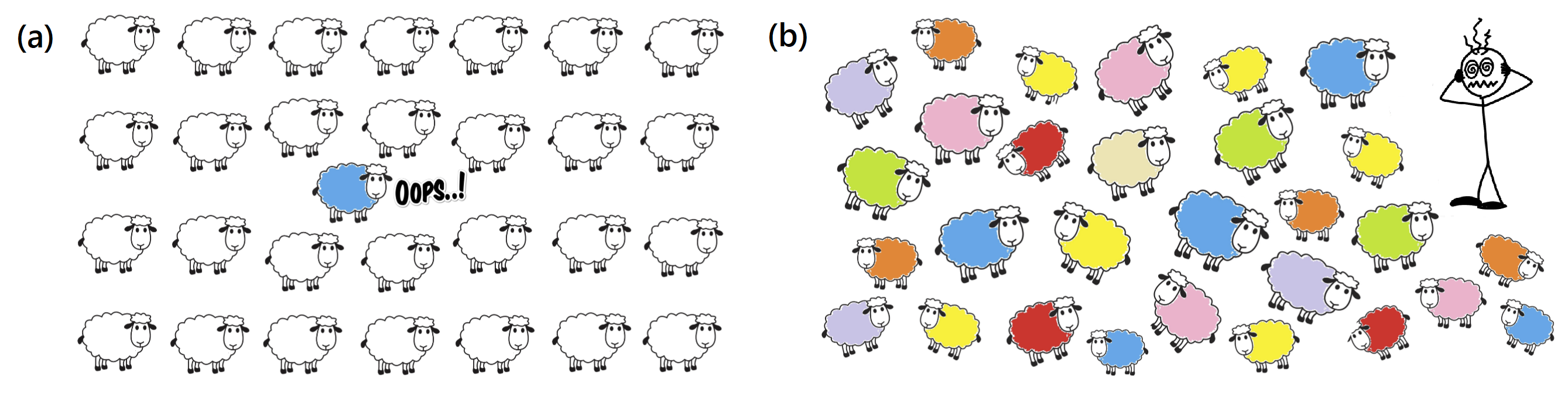}
\caption{\textbf{Finding defects in disorder systems is a difficult task}  \textbf{(a)} It is rather simple to find a defect (blue sheep) in a periodic ordered structure (white sheeps), as in crystalline matter with long-range order. \textbf{(b)} It is (almost) impossible to define a defect in a disordered structure like this herd of colored sheeps, or like in a real glass.}
\label{fig:figure1}
\end{figure}

In a crystalline solid, atoms are not randomly arranged. On the contrary, they sit in preferred positions, forming a periodic and ordered structure as depicted by the white sheeps in Fig.\ref{fig:figure1}\color{blue}a\color{black}. In technical words, this is known as long-range order, and it is the key behind several of the physical properties of crystals, from their rigidity to the propagation of sound and heat transport. Whenever such an organized configuration exists, it is almost immediate to find out a particle which does not follow the rules and breaks the order. In physics, we call such a rebel, like the blue sheep in Fig.\ref{fig:figure1}\color{blue}a\color{black}, a defect.

If all solids were similarly well-behaved, the life of a physicist would be rather boring. Nevertheless, many solids in Nature are amorphous, and they do not respect the ordered and periodic atomic arrangement described above. Glasses are the most famous example of that sort. The structure of a glass is disordered and more akin to the colorful and heterogeneous herd of sheeps shown in Fig.\ref{fig:figure1}\color{blue}b\color{black}. As a consequence, identifying a defect therein appears as an almost impossible task, if not an ill-defined concept altogether.

The reason why we should care about finding defects in that mess is very practical. In crystalline solids, defects play a fundamental role in predicting mechanical failure and the onset of plasticity \cite{doi:10.1098/rspa.1934.0106}, the irreversible deformations which bring to the breakdown of the elastic response. In other words, they are critical to connect structure with dynamics, and to predict when and where a certain material will break. A common idea is that, even in glasses, plastic deformations take place in soft spots with abnormally low elastic constants and increased mobility, analogous to dislocations in crystalline systems. The remaining question is how to locate those weak zones from structural information, or even better how to relate them to the presence of defects which potentially control the plastic flow. In order to understand the difficulty behind this task, we need to dive into the mathematical tools that physicists use to define and quantify order and disorder.\\

\begin{figure}[h]\centering
\includegraphics[width=\linewidth]{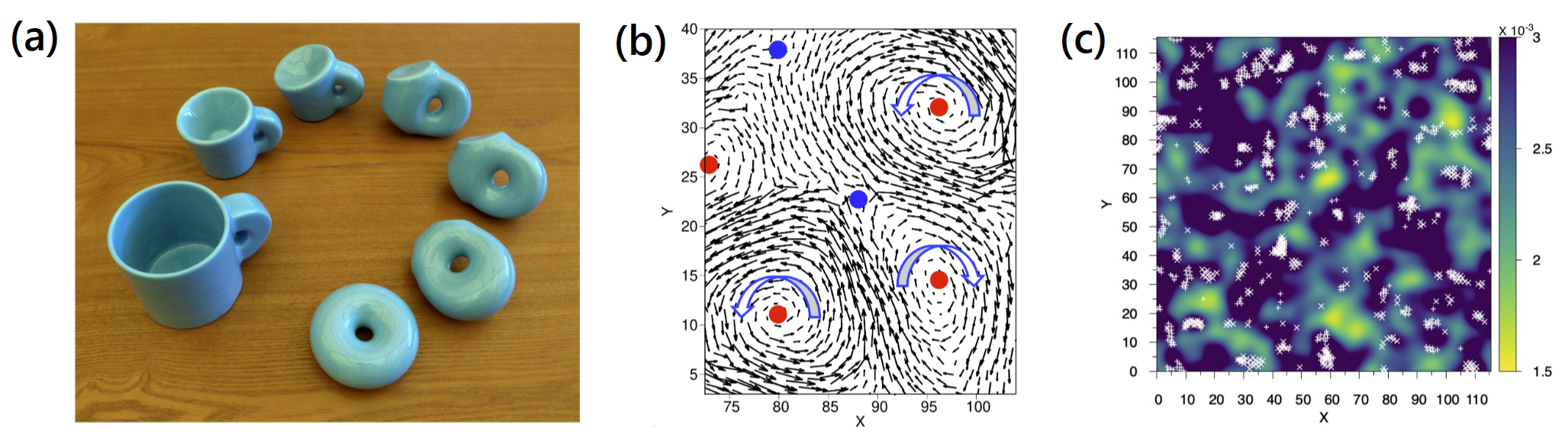}
\caption{\textbf{The revealed link between topology and plasticity} \textbf{(a)} A doughnut can be continuously deformed into a coffee mug, hence the two objects are topologically equivalent: they both have one hole. Ceramic model by Keenan Crane and Henry Segerman. \textbf{(b)} The identification of topological defects with positive (red) and negative (blue) charges in the normal modes eigenvectors in the work by Wu et al. \cite{wu2022topology}. \textbf{(c)} The correlation between the plastic events (white crosses) and the density of topological defects with negative charge (color map) presented in the work by Wu et al. \cite{wu2022topology}.}
\label{fig:figure2}
\end{figure}

Topological defects appear in many disparate areas of physics from cosmic strings in the universe, to vortices in superfluids and even in the patterns of human fingerprints. They represent a beautiful bridge between physics and a branch of mathematics known as topology. Given an ordered medium, like the sheeps in Fig.\ref{fig:figure1}\color{blue}a\color{black}, its configuration can be mathematically described by an order parameter, which defines a map between static structures in the real space and the allowed ground states in the energy space of the system. Defects are singularities of this order parameter field which cannot be removed by a smooth deformation, without gluing and tearing parts. They are characterized by a winding number, which corresponds to the total angle through which the order parameter rotates as one surrounds the defect with a closed loop. This number can be thought as a (topological) charge for the defects, by analogy with point-like charged particles in electromagnetism. More mathematically, defects can be classified by their homotopy group, which measures the topological properties of a certain manifold, such as the number of holes in it. A practical application of these concepts leads mathematicians, and quirky potters (see Fig.\ref{fig:figure2}\color{blue}a\color{black}), to conclude that a coffee mug and a doughnut are equivalent, since they share the same type of defect. 

To avoid confusion, let us clarify that with the word topological we do not mean defects in coordination or in the number of neighbors, as sometimes used in the context of disordered materials. In the mathematical terms outlined above, those are not topological since they can be removed by a continuous transformation. Beside all the jargon, the main point is that the definition of defects necessitates the existence of an unbroken subset of symmetries that leaves the ground state of the system invariant. For crystalline solids, these are just the rigid translations of the ordered lattice. For amorphous systems, there is no remaining isotropy group. The whole shebang collapses from the start. To iterate the concept, a certain degree of order is needed to define disorder.

The search for topological defects in amorphous solids has a long and controversial history. The idea that dislocations lines could exist in glasses is almost $50$ years old \cite{gilman1973flow}, and has been extensively scrutinized and questioned using theoretical arguments and numerical simulations \cite{PhysRevLett.43.1517,doi:10.1080/01418618108235816,doi:10.1080/01418618008243894}. Studies have shown that the properties of amorphous systems are undeniably structure sensitive and that regions of high stress and low symmetry, resembling dislocation cores in crystals, can be identified in glasses \cite{doi:10.1080/01418618008243894}.

The conviction that plasticity in glasses could still be related to structure led to the introduction of a plethora of structural, thermodynamic and mechanical indicators, including the local shear modulus, energetically favoured regions, linear and non-linear vibrational modes, local thermal energy and more abstract measures of softness. A throughout study \cite{PhysRevMaterials.4.113609} concluded that most of these indicators are excellent at locating plastic events over short strain scales, but they do not provide a first principles understanding of plasticity, as in crystals. On the other hand, several successful theories such as shear transformation zones \cite{PhysRevE.57.7192}, and elasto-plastic models \cite{RevModPhys.90.045006}, have postulated the existence of such defects without any precise definition.

Against this background, one could just rely on these successful but phenomenological structural indicators or search for defects in glasses beyond their real-space structure. This second choice is what all these new developments are about. The first idea in Ref.~\cite{PhysRevLett.127.015501} was to hunt for defects in the dynamical displacement field rather than in the static structure, and look for singularities upon deforming the system. The inspiration came by thinking about the incompatibility of the deformation, which naturally arises because of non-affinity, and which can be related using mathematical objects known as higher-form symmetries to a strain-free formulation of elasticity \cite{PhysRevE.105.024602}. The results of \cite{PhysRevLett.127.015501} indicated that standard topological concepts applied to the dynamical displacement field allow for a precise identification of defects, which well correlate with the major plastic events and successfully predict the location of the global yielding instability.

Still, the study in \cite{PhysRevLett.127.015501} gave up on probing the static structure. In other words, the problem of relating structural defects to dynamics was bypassed focusing only on the dynamics itself. That was still unsatisfactory, until Wu and colleagues' recent paper, published in Nature Communications \cite{wu2022topology}, came out and provided a potential breakthrough in this story. 

Wu and colleagues \cite{wu2022topology} discovered that the long sought structural defects in glasses were hiding in the topology of the vibrational eigenmodes. Differently from the original idea of \cite{PhysRevLett.127.015501}, that is still a property of the static and undeformed configuration, but not of the real-space structure. Wu and colleagues noticed that the spatial distribution of the eigenvectors display a collection of whirls and curls and eye-visible vortex structures (see Fig.\ref{fig:figure2}\color{blue}b\color{black}), with manifest singular behavior. By surrounding these defects with closed loops, and measuring the angular decifit of the vectorial field around them, they were able to obtain the corresponding topological charges and identify positive and negative defects. Positive ones are just perfect vortices, like those in the swirling water of your bathtub. Negative ones correspond to frustrated interfaces with a saddle shape. Using advanced statistical methods, Wu and colleagues were able to show a solid correlation between the density of negative defects and the location of the plastic events, defined using the widely accepted concept of non-affine displacement \cite{PhysRevE.57.7192}. In Fig.\ref{fig:figure2}\color{blue}c\color{black}, visual evidence of this result is presented, where the plastic events, shown with white crosses, nicely correlate with the darker area with higher density of negative defects. 

To make it short, Wu and colleagues \cite{wu2022topology} managed for the first time to identify topological defects in the static structure of glasses and to provide a direct link between the properties of glasses before deformation and the plastic behavior during it. As for the case of dislocations in crystalline solids, this realizes a connection between structure and dynamics in amorphous systems which goes far beyond all the phenomenological structural indicators considered before and highlights the role of topology and geometry in the context of disordered systems and their plastic behavior. 

The revealed topological information displays striking similarities with the quadrupolar Eshelby-like structures believed to be fundamental for the plasticity of amorphous solids \cite{PhysRevE.104.024904}, and the vortex-like formations possibly constituting the shear transformation zones in glasses \cite{PhysRevLett.119.195503}. Finally, these defects could be formally related to geometric charges in the metric formulation of elasticity \cite{Moshe}, and to magnetic currents in the gauge theories for emergent elasticity in granular matter \cite{PhysRevE.106.065004}.

In summary, the topological defects discovered by Wu and colleagues could play a pivotal role in our understanding of glasses, from the boson peak feud, to the nature of the glass transition as a topological phase transition, up to the fundamental origin of yielding. We just need to sit with a coffee mug and a doughnut and wait to see how topology will help us to make order out of disorder.

\section*{Acknowledgements}
I would like to thank Alessio Zaccone, Michael Landry, Yuliang Jin, Deng Pang, Wanzhou Zhang, Yunjiang Wang and Jie Zhang for discussions about symmetries, topological defects and plasticity in amorphous systems, and for useful comments about a preliminary version of this manuscript.
I acknowledge the support of the Shanghai Municipal Science and Technology Major Project (Grant No.2019SHZDZX01) and the sponsorship from the Yangyang Development Fund. 

\section*{Author contributions}
M.B. did everything.

\section*{Competing interests}
The author declares no competing interests.

\end{document}